\pdfoutput=1
\documentclass[%
 reprint,
amsmath,amssymb,
aps,
prl
]{revtex4-1}

\usepackage{graphicx}
\usepackage{dcolumn}
\usepackage{bm}

\usepackage{amsmath,amssymb,amsbsy,amstext,amsthm,simplewick,amsfonts,graphicx}
\usepackage{mathrsfs}
\usepackage{graphicx}
\usepackage{wrapfig}
\usepackage{upgreek}
\usepackage{bm} 
\usepackage{framed}
\usepackage{bbm}
\usepackage{textcomp}
\usepackage{adjustbox}
\usepackage{makecell}
\usepackage{tcolorbox}
\usepackage{empheq}
\usepackage[normalem]{ulem}
\usepackage{enumitem}
\usepackage{braket} 
\usepackage{array}
\usepackage{dsfont}
\usepackage{ulem}
\usepackage{xcolor}
\usepackage{caption}
\usepackage{subcaption}
\usepackage{tocloft}
\usepackage{tikz}
\usepackage{xcolor}
\usepackage{leftindex}
\usetikzlibrary{patterns}
\usetikzlibrary{patterns}
\usetikzlibrary{decorations.markings}
\usetikzlibrary{decorations.pathmorphing}
\usetikzlibrary{shapes.misc}
\usetikzlibrary{arrows}
\usetikzlibrary{shapes.geometric}
\usetikzlibrary{plotmarks}
\usepackage{hyperref}
\usepackage{titlesec}
\hypersetup{colorlinks=true,
	breaklinks=true,
	pdfstartview=Fit,
	linkcolor=blue2,
	citecolor=red2,
	urlcolor=blue
}

\definecolor{red2}{RGB}{214, 39, 40}
\definecolor{green2}{RGB}{0,170,0}
\definecolor{blue2}{RGB}{0,100,200}
\definecolor{magenta2}{RGB}{191,64,191}
\definecolor{purple2}{RGB}{112,48,160}
\definecolor{orange2}{RGB}{255,192,0}

\usepackage{graphicx,scalerel}

\graphicspath{{Fig/}}

\def\Re{\mathrm{Re}\,}
\def\Im{\mathrm{Im}\,}

\definecolor{rosewood}{rgb}{0.4, 0.0, 0.04}
\definecolor{pyblue}{RGB}{31, 119, 180}
\definecolor{pyorange}{RGB}{255, 127, 14}
\definecolor{pygreen}{RGB}{44, 160, 44}
\definecolor{pyred}{RGB}{214, 39, 40}
\definecolor{pypurple}{RGB}{148, 103, 189}
\definecolor{lightgray}{gray}{0.9}

\begin{document}

\setlength{\parskip}{0.1\baselineskip}
\preprint{APS/123-QED}

\titlespacing*{\section}
{0pt}   
{8pt}   
{4pt}   

\title{\textbf{Conformal Anomalies and Generalised Energy Conservation\\ in De Sitter Spacetime}}

\author{Zong-Zhe Du}
\email{z.du1@imperial.ac.uk}
 \affiliation{Abdus Salam Centre for Theoretical Physics, Imperial College, London SW7 2AZ, United Kingdom}

\date{\today}

\vspace{20mm}

\begin{abstract}
 We show that for bulk discrete series fields whose dual operators have integer conformal dimensions, the Generalised Energy Conservation (GEC) condition for de Sitter (dS) scattering amplitudes defined in even-dimensional extended Poincar\'e spacetime corresponds precisely to the absence of boundary conformal anomalies, including beta-function contributions, at the level of tree-level contact interactions. In contrast, holographic consistency in AdS/CFT enforces only the locality of the conformal anomaly.
 We find novel, unique four-point interactions involving three identical conformally coupled scalars and one discrete series scalar by performing both GEC and holographic bootstraps. This demonstrates that the solutions to GEC form a measure-zero subset of the holographically consistent solutions. Our holographic analysis agrees precisely with the expected renormalisation conditions for conformal field theory correlators. In addition, the constraint for a particular four-point configuration may be relevant for testing holographic consistency of scale-separated non-supersymmetric AdS$_4$ vacua.
        \end{abstract}

\maketitle
	\setcounter{page}{1} 
	\thispagestyle{empty}
\noindent

\section{Generalised energy conservation and holographic consistency}

Generalised Energy Conservation is the condition that the dS scattering amplitude defined in extended Poincar\'e dS spacetime has support only when the total energy is conserved \cite{bsyw-czt5},
\begin{align}
    \mathcal{A}_{k_T\neq 0} = 0,
\end{align}
where $k_T = \sum k_{\rm in} - \sum k_{\rm out}$, and the $k_a$ are the magnitudes of the flat-slicing momenta, $k_a \equiv \sqrt{g_{\alpha\beta}k_a^{\alpha}k_a^{\beta}}$, which we refer to as energies when $g_{\alpha\beta} = \delta_{\alpha\beta}$. As proved and illustrated by non-trivial examples in \cite{bsyw-czt5}, the scattering amplitude is invariant under perturbative field redefinitions. When imposed on discrete series scalars, GEC naturally reproduces the known exceptional scalar interactions in $D=4$: the $\Delta=4$ Dirac-Born-Infeld theory \cite{Goon:2011qf,Bonifacio:2021mrf} and the $\Delta=5$ special Galileon \cite{Bonifacio:2018zex,Bonifacio:2021mrf}. Although higher-$\Delta$ theories are generally unknown, GEC is powerful enough to determine their unique four-point scalar self-interactions up to a given derivative order. This selection rule is reminiscent of holographic consistency in AdS/CFT, according to which the bulk coupling responsible for a divergent correlator vanishes. Specifically, all (super)extremal correlators satisfying $\Delta_i-\Delta_j-\Delta_k=2n$, with $n$ a non-negative integer, are divergent; the corresponding bulk coupling must therefore vanish, while the correlator remains finite. This conjecture has been extensively verified in several settings, including $\mathcal{N}=2$ supergravity on $AdS_5\times S^5$ \cite{DHoker:1999jke,Arutyunov:1999fb} and massive type-IIA superstring flux compactifications to AdS \cite{Apers:2022tfm,Bobev:2025yxp}. Originally formulated in position space, the conjecture was recast elegantly in momentum space in \cite{Bzowski:2015pba}. The underlying argument is that a (super)extremal correlator cannot be renormalised by a boundary-local counterterm, assuming holographic renormalisability, the corresponding bulk coupling must therefore vanish. See \cite{Bzowski:2019kwd} for a discussion of higher-point functions, to which we return below. See \cite{Castro:2024cmf} for a caveat \footnote{ These bulk vertices can be renormalised by introducing counterterms local in canonical momentum, which are generically non-local in source, in this work we exclude such scenario.}. Both GEC and holographic consistency thus impose strong constraints on bulk couplings. In this work, we establish a connection between them.

\begin{figure}
    \centering

\tikzset{every picture/.style={line width=0.75pt}} 

\begin{tikzpicture}[x=0.75pt,y=0.75pt,yscale=-0.65,xscale=0.65]

\draw   (222,50) -- (423,50) -- (423,251) -- (222,251) -- cycle ;
\draw [color={rgb, 255:red, 74; green, 144; blue, 226 }  ,draw opacity=1 ][line width=2.25]    (423,50) -- (222,251) ;
\draw [color={rgb, 255:red, 208; green, 2; blue, 27 }  ,draw opacity=1 ][line width=2.25]    (222,50) -- (423,50) ;
\draw [color={rgb, 255:red, 208; green, 2; blue, 27 }  ,draw opacity=1 ][line width=2.25]    (222,251) -- (423,251) ;
\draw [line width=2.25]    (288.5,64) -- (288.5,34) ;
\draw [shift={(288.5,29)}, rotate = 90] [fill={rgb, 255:red, 0; green, 0; blue, 0 }  ][line width=0.08]  [draw opacity=0] (14.29,-6.86) -- (0,0) -- (14.29,6.86) -- cycle    ;
\draw [line width=2.25]    (345.5,64) -- (345.5,34) ;
\draw [shift={(345.5,29)}, rotate = 90] [fill={rgb, 255:red, 0; green, 0; blue, 0 }  ][line width=0.08]  [draw opacity=0] (14.29,-6.86) -- (0,0) -- (14.29,6.86) -- cycle    ;
\draw [line width=2.25]    (292.5,265) -- (292.5,235) ;
\draw [shift={(292.5,230)}, rotate = 90] [fill={rgb, 255:red, 0; green, 0; blue, 0 }  ][line width=0.08]  [draw opacity=0] (14.29,-6.86) -- (0,0) -- (14.29,6.86) -- cycle    ;
\draw [line width=2.25]    (346.5,265) -- (346.5,235) ;
\draw [shift={(346.5,230)}, rotate = 90] [fill={rgb, 255:red, 0; green, 0; blue, 0 }  ][line width=0.08]  [draw opacity=0] (14.29,-6.86) -- (0,0) -- (14.29,6.86) -- cycle    ;

\draw (169,41.4) node [anchor=north west][inner sep=0.75pt]  [color={rgb, 255:red, 208; green, 2; blue, 27 }  ,opacity=1 ,xscale=0.75,yscale=0.75 ]  {$\eta =0_{-}$};
\draw (168,243.4) node [anchor=north west][inner sep=0.75pt]  [color={rgb, 255:red, 208; green, 2; blue, 27 }  ,opacity=1 ,xscale=0.75,yscale=0.75 ]  {$\eta =0_{+}$};
\draw (310,85.4) node [anchor=north west][inner sep=0.75pt]  [color={rgb, 255:red, 74; green, 144; blue, 226 }  ,opacity=1 ,xscale=0.75,yscale=0.75 ]  {$\eta =-\infty $};
\draw (359,112.4) node [anchor=north west][inner sep=0.75pt]  [color={rgb, 255:red, 74; green, 144; blue, 226 }  ,opacity=1 ,xscale=0.75,yscale=0.75 ]  {$\eta =+\infty $};
\draw (450,44) node [anchor=north west][inner sep=0.75pt]  [color={rgb, 255:red, 74; green, 144; blue, 226 }  ,opacity=1 ,xscale=0.75,yscale=0.75 ] [align=left] {Horizon};
\draw (74,137) node [anchor=north west][inner sep=0.75pt]  [color={rgb, 255:red, 208; green, 2; blue, 27 }  ,opacity=1 ,xscale=0.75,yscale=0.75 ] [align=left] {Conformal Boundary};
\draw (304,173.4) node [anchor=north west][inner sep=0.75pt]    {$\langle out|$};
\draw (284,136.4) node [anchor=north west][inner sep=0.75pt]    {$|in\rangle$};
\draw    (150.54,133) -- (187.19,63) ;
\draw    (149.81,158) -- (187.46,239) ;
\draw    (460.61,65) -- (404.23,108) ;
\draw    (447,61.3) -- (371,83.61) ;

\end{tikzpicture}

    \caption{A Penrose-diagrammatic representation of the dS scattering amplitude. Particles are prepared at the past horizon, cross the boundary along a complex deformation (represented by the bold arrows), and end at the future horizon. Boundary regularisation is generally required for IR-divergent theories. The resulting scattering amplitude is invariant under field redefinitions and conserves energy only for a special class of theories, some of which coincide with exceptional EFTs \cite{bsyw-czt5}.
    }
    \label{fig:dSScattering}
\end{figure}
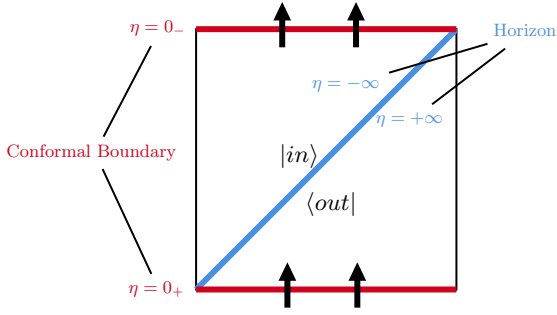

\begin{figure}
    \centering

\tikzset{every picture/.style={line width=0.75pt}} 

\begin{tikzpicture}[x=0.75pt,y=0.75pt,yscale=-0.65,xscale=0.65]

\draw   (124,48) -- (325,48) -- (325,249) -- (124,249) -- cycle ;
\draw [color={rgb, 255:red, 74; green, 144; blue, 226 }  ,draw opacity=1 ][line width=2.25]    (323.5,149) -- (124,249) ;
\draw [color={rgb, 255:red, 208; green, 2; blue, 27 }  ,draw opacity=1 ][line width=2.25]    (325,249) -- (325,48) ;
\draw [color={rgb, 255:red, 208; green, 2; blue, 27 }  ,draw opacity=1 ][line width=2.25]    (124,249) -- (124,48) ;
\draw [color={rgb, 255:red, 74; green, 144; blue, 226 }  ,draw opacity=1 ][line width=2.25]    (323.5,149) -- (124,48) ;
\draw [line width=2.25]    (308.5,124) -- (344.5,123.12) ;
\draw [shift={(349.5,123)}, rotate = 178.6] [fill={rgb, 255:red, 0; green, 0; blue, 0 }  ][line width=0.08]  [draw opacity=0] (14.29,-6.86) -- (0,0) -- (14.29,6.86) -- cycle    ;
\draw [line width=2.25]    (106.5,122) -- (142.5,121.12) ;
\draw [shift={(147.5,121)}, rotate = 178.6] [fill={rgb, 255:red, 0; green, 0; blue, 0 }  ][line width=0.08]  [draw opacity=0] (14.29,-6.86) -- (0,0) -- (14.29,6.86) -- cycle    ;
\draw [line width=2.25]    (108.5,162) -- (144.5,161.12) ;
\draw [shift={(149.5,161)}, rotate = 178.6] [fill={rgb, 255:red, 0; green, 0; blue, 0 }  ][line width=0.08]  [draw opacity=0] (14.29,-6.86) -- (0,0) -- (14.29,6.86) -- cycle    ;
\draw [line width=2.25]    (307.5,170) -- (343.5,169.12) ;
\draw [shift={(348.5,169)}, rotate = 178.6] [fill={rgb, 255:red, 0; green, 0; blue, 0 }  ][line width=0.08]  [draw opacity=0] (14.29,-6.86) -- (0,0) -- (14.29,6.86) -- cycle    ;

\draw (64,231.4) node [anchor=north west][inner sep=0.75pt]  [color={rgb, 255:red, 208; green, 2; blue, 27 }  ,opacity=1 ,xscale=0.75,yscale=0.75 ]  {$z=0_{+}$};
\draw (344,226.4) node [anchor=north west][inner sep=0.75pt]  [color={rgb, 255:red, 208; green, 2; blue, 27 }  ,opacity=1 ,xscale=0.75,yscale=0.75 ]  {$z=0_{-}$};
\draw (178,113.4) node [anchor=north west][inner sep=0.75pt]  [color={rgb, 255:red, 74; green, 144; blue, 226 }  ,opacity=1 ,xscale=0.75,yscale=0.75 ]  {$z=+\infty $};
\draw (221,76.4) node [anchor=north west][inner sep=0.75pt]  [color={rgb, 255:red, 74; green, 144; blue, 226 }  ,opacity=1 ,xscale=0.75,yscale=0.75 ]  {$z=-\infty $};
\draw (203,5) node [anchor=north west][inner sep=0.75pt]  [color={rgb, 255:red, 74; green, 144; blue, 226 }  ,opacity=1 ,xscale=0.75,yscale=0.75 ] [align=left] {Horizon};
\draw (161,300) node [anchor=north west][inner sep=0.75pt]  [color={rgb, 255:red, 208; green, 2; blue, 27 }  ,opacity=1 ,xscale=0.75,yscale=0.75 ] [align=left] {Conformal Boundary};
\draw (267,98.4) node [anchor=north west][inner sep=0.75pt]    {$|in\rangle$};
\draw (240,133.4) node [anchor=north west][inner sep=0.75pt]    {$\langle out| $};
\draw    (205,296) -- (115.54,253) ;
\draw    (254.3,296) -- (343.77,248) ;
\draw    (233.46,26) -- (246.18,72) ;
\draw    (227.33,26) -- (209.57,109) ;

\end{tikzpicture}

    \caption{Setup for the scattering amplitude in AdS. The amplitude is invariant under field redefinitions, and energy is always conserved; therefore, no instability can be inferred from it.
    }
    \label{fig:AdSScattering}
\end{figure}
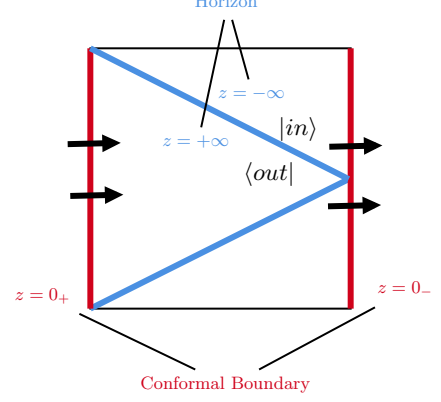

We next motivate the study of dS scattering amplitudes. It is well known that boundary correlators are not invariant under bulk field redefinitions: a perturbative field redefinition generates new boundary sources that modify the correlator by local counterterms \cite{Arutyunov:2000a}. However, verifying holographic consistency requires a sequence of complicated field redefinitions to reveal the vanishing bulk coupling \cite{Arutyunov:1999fb}. A direct computation is therefore impractical because boundary correlators depend sensitively on the field basis. Therefore, a field-basis-independent observable is yearned. The dS scattering amplitude provides such a meta-observable on extended Poincar\'e dS spacetime \cite{Donath:2024utn}:
\begin{align}
    ds^2 = \frac{-d\eta^2 + g_{\alpha\beta}dx^\alpha dx^\beta}{H^2 \eta^2},\; \eta\in (-\infty, -\epsilon)\cup \mathcal{C}_\epsilon \cup(\epsilon,+\infty),\label{metric}
\end{align}
where $\mathcal{C}_{\epsilon}$ is a complex deformation around the conformal boundary, elucidated in the Supplemental Material. The \textit{in} and \textit{out} states are located at the horizons $\eta=-\infty$ and $\eta=+\infty$, respectively, while the two patches are glued near the conformal boundary through the complex deformation. The \textit{in} state is prepared at the past horizon, evolves across the boundary along the deformed contour, and becomes the \textit{out} state at the future horizon, see Figure \ref{fig:dSScattering}.
The scattering amplitude realises the local dS isometries and is invariant under changes of field basis, as proved using LSZ reduction and demonstrated through examples in \cite{bsyw-czt5}. In particular, the scattering amplitude is well defined at tree level without renormalisation; only the time integral must be regularised by the small deformation of the integration contour around the infrared pole at the conformal boundary. The S-matrix and scattering amplitudes 
can be parametrised as \cite{bsyw-czt5}
\begin{align}
\braket{out|in}&= \braket{\beta|\text{exp}\biggl(-i\int _{-\infty}^{+\infty} \mathcal{H}_{\text{int}} d\eta \biggr)|\alpha} \nonumber\\
&\equiv 1 + i(2\pi)^{D} \mathcal{A}\;\delta^{D-1}({\Vec{k}_{out}-\Vec{k}_{in}})\\
\text{where}\;\;\mathcal{A}_{\mp} &= \mp\mathcal{A}_{k_T\neq 0} \theta(\pm k_T) + \sum_{m} \mathcal{A}_{m} \partial_{k_T}^m \delta(k_T),
\end{align}
where $D$ is the dS spacetime dimension, and the $\pm$ sign denotes the scheme dependence, namely the direction of the contour deformation described in the Supplemental Material. In forthcoming work, we argue that this structure holds for arbitrary tree-level amplitudes and mass spectra. The presence of a $\theta(-k_T)$ term signals eternal energy creation. When integrated over phase space, this term produces a UV divergence, suggesting a possible vacuum instability. We therefore conjecture that GEC is a necessary criterion for dS vacua in quantum gravity \cite{bsyw-czt5}, in agreement with dS/CFT examples \cite{Anninos:2014lwa}, as explained below. The gluing of $\eta=0^-$ and $\eta=0^+$ is also motivated by a Euclidean path-integral construction \cite{Witten:2024upt,Anninos:2026hia}, in which Euclidean dS, $S^4$, is covered by two hemispheres whose $S^3$ boundaries are identified. The construction applies equally to AdS, see Figure \ref{fig:AdSScattering}.
From the scattering-amplitude perspective, however, no instability arises because energy is always conserved in AdS. Eternal momentum creation or annihilation in the AdS radial direction does not introduce an inconsistency because it is constrained by energy conservation. Instead, holographic consistency serves as a guiding principle for constructing effective field theories (EFTs) in AdS spacetime \cite{El-Showk:2011yvt}. Here, we relate this (A)dS holographic consistency condition to energy conservation in dS.

Veering back to holographic correlators, operators of integer conformal dimension in an odd-dimensional boundary CFT generally produce a logarithmic divergence that cannot be removed by a perturbative bulk field redefinition, or equivalently by a renormalisation of the source \cite{Petkou:1999fv}:
\begin{align}
    \psi_n\equiv\braket{\mathcal{O}_1(k_1)\ldots\mathcal{O}_n(k_n)} \supset -a\;\text{log}\frac{k_T}{\mu},
\end{align}
where the $\mathcal{O}_i$ are boundary operators and $\mu$ is the renormalisation scale. The scale dependence signals a conformal anomaly or beta function contribution, with $a$ identified as the anomaly coefficient, it is also degenerate with the corresponding counterterm. This interchangeability between the counterterm and the anomaly underlies the scheme dependence of holographic renormalisation \cite{Bzowski:2015pba}. Throughout this work, we use ``anomaly'' collectively for all sources of scale dependence, including both the conformal anomaly and beta function contributions. Here, $k_T$ denotes the sum of the AdS radial momenta $k_a \equiv \sqrt{-g_{\alpha\beta}k_a^{\alpha}k_a^{\beta}}$, where $g_{\alpha\beta}$ is again the boundary metric. In Lorentzian AdS, $g_{\alpha\beta}= \eta_{\alpha\beta}$, the minus sign arises from the continuation  $\eta = - iz$. Henceforth, we write $k_a\equiv\sqrt{\pm g_{\alpha\beta}k_a^{\alpha}k_a^{\beta}}$, where the plus and minus signs correspond to dS and AdS, respectively, and $g_{\alpha\beta}$ denotes the corresponding flat boundary metric. The scheme dependence of the scattering amplitude consists of the two discrete choices $\pm k_T$, whereas that of the correlator involves the continuous scale $\mu$. This distinction underlies the mathematical rigidity of the scattering amplitude. At the level of tree-level contact interactions for discrete series fields in even $D$, the holographic correlator and scattering amplitude reduce to the two integrals
\begin{widetext}
    \begin{align}
\psi_n &\sim  i\int_{\textcolor{pyred}{0}}^{\textcolor{pyred}{\infty}} \frac{d z}{z^D} F(ikz)  e^{-ik_T z}
\stackrel{\text{renormalisation}}{\supset} -a\;\text{log}\frac{k_T}{\mu},\\
    2\pi i \mathcal{A}(k_1,\ldots,k_n) &\sim i\int_{\textcolor{pyred}{-\infty}}^{\textcolor{pyred}{\infty}} \frac{d\eta}{\eta^D} F(ik\eta) e^{-ik_T \eta}
    \stackrel{\text{regularisation}}{\supset} \mp 2\pi  i\mathcal{A}_{k_T\neq 0} \theta(\pm k_T),
\end{align}
\end{widetext}
where the denominator comes from the spacetime measure. The generic elementary function $F(x)$ \footnote{In even $D$, the mode functions \eqref{ModeFunc} reduce to elementary, meromorphic functions.} captures the details of the contact interaction and can be constructed from the mode functions
\begin{align}\label{ModeFunc}
f_k(\eta)&=\frac{\sqrt{\pi}H^{\frac{D-2}{2}}}{2} (-\eta)^{\frac{D-1}{2}} e^{i\frac{\pi}{2}(\nu - 1/2)} H_{\nu}^{(1)}(-k\eta),\\
   \phi_k(\eta) &= k^{\nu} f_k(\eta),\label{BulktoBoundary}\qquad\nu= \Delta - \frac{D-1}{2},
\end{align}
where $\phi_k(\eta)$ is the bulk-to-boundary propagator. For half-integer $\nu$, the propagators reduce to meromorphic elementary functions. The integrands for the amplitude and the correlator are therefore identical. For the correlator, one integrates over a single Poincar\'e patch, whereas for the amplitude one integrates over both Poincar\'e patches. Importantly, amplitudes and correlators can each be defined in either dS or AdS, although their physical interpretations differ. The comparison between dS and AdS is justified if perturbation theory is valid in both vacua: one makes the substitution $H^2=-1/R^2,\;\;\eta = - i z$ while keeping the conformal dimension $\Delta$ fixed. Consequently, $m^2$ changes sign:
\begin{align}
    m^2 = - \Delta(\Delta-d)H^2\leftrightarrow m^2 = \Delta(\Delta-d)\frac{1}{R^2}.
\end{align}
The major difference lies in the physical regime of $k_a$. We have singled out a universal phase factor $e^{-ik_T\eta}$ to make the convergence of the amplitude integral as $|\eta|\rightarrow\infty$ manifest. Note that the normalisation used here differs from that of \cite{bsyw-czt5}, see footnote \footnote{Our normalisation of one-particle states is
\begin{align}
    \ket{\vec{k}}=k^{\nu}a^\dag_{\vec{k}}\ket{0},
\end{align}
so that the two-point S-matrix is conformally invariant:
\begin{align}
    \braket{\vec{k}_1|\vec{k}_2} = (2\pi)^{D-1} k^{2\nu}\delta^{D-1}(\vec{k}_1-\vec{k}_2),
\end{align}
which reproduces the well-cherished momentum-space two-point function of a CFT for those operator dimensions.}. In a forthcoming work, we demonstrate that the S-matrix can be interpreted as a full-fledged holographic correlator.

By the residue theorem, deforming the integration contour around the infrared pole at $\eta=0$ produces the non-energy-conserving term in the scattering amplitude, as detailed in the Supplemental Material and in \cite{bsyw-czt5}:
\begin{align}
    &\int_{-\infty}^{+\infty}\frac{d\eta}{\eta^D} F(ik\eta) e^{-ik_T\eta} \nonumber\\
    &\supset \mp 2\pi i \text{Res}_{\eta=0}\;\biggl[\frac{1}{\eta^D} F(ik\eta)e^{-ik_T\eta}\biggr] \theta(\pm k_T).
\end{align}
The non-energy conserving amplitude is therefore
\begin{align}
    \mathcal{A}_{k_T\neq 0} = i \text{Res}_{\eta=0}\;\biggl[\frac{1}{\eta^D} F(ik\eta)e^{-ik_T\eta}\biggr].
\end{align}
The correlator likewise suffers from an IR divergence at the conformal boundary. In this work, we adopt a finite-cutoff regularisation scheme by placing the cutoff at $z=\epsilon$. The regulated correlator is
\begin{align}
  \psi_n(\epsilon) = i\int_{\epsilon}^{+\infty} \frac{dz}{z^D} F(ikz)e^{-ik_T z},\label{Correlator}
\end{align}
which admits the following asymptotic expansion for small $\epsilon$:
\begin{align}
     \psi_n(\epsilon) = \sum_{n\leq 0} f_n\epsilon^n - a\; \text{log}\epsilon +\mathcal{O}(\epsilon).
\end{align}
Terms with negative $n$ can be removed completely by boundary-local counterterms \cite{Skenderis:2002wp,Petkou:1999fv,Nojiri:2000kz,Henningson:1998gx}. The logarithmic term induces the conformal anomaly $a$, and its divergence is cancelled by the counterterm
\begin{align}
    \psi_{\text{ct}}= a\; \text{log} (\mu \epsilon),
\end{align}
assuming its existence. The anomaly coefficient $a$ is therefore the coefficient of the logarithmic divergence and can be extracted from
\begin{align}
    \partial_\epsilon \psi_n(\epsilon) = \sum_{n\leq-2} g_n \epsilon^n - \frac{a}{\epsilon} + \mathcal{O}(1).
\end{align}
Inserting the correlator \eqref{Correlator} into this expression and applying the fundamental theorem of calculus gives
\begin{align}
    \partial_\epsilon \psi_n(\epsilon) = - i \frac{1}{\epsilon^{D}} F(ik \epsilon) e^{- i k_T \epsilon}.
\end{align}
The coefficient of $\frac{1}{\epsilon}$ in the Laurent expansion is therefore
\begin{align}
    a = i\text{Res}_{z=0}\biggl[ \frac{1}{z^D}F(ikz) e^{-i k_T z}\biggr].
\end{align}
Therefore, the non-energy conserving scattering amplitude can be precisely identified with the anomaly $a$ for contact interactions when combining the above equations:
\begin{align}\label{Main}\boxed{
  a = \mu \frac{\partial}{\partial \mu} \psi_n = \mathcal{A}_{k_T\neq 0},}
\end{align}
The factor of $2\pi$ cancels in our convention. This is mildly surprising, as the cancellation is not entirely by construction. Because the scattering amplitude is invariant under field redefinitions \cite{bsyw-czt5}, \eqref{Main} can be regarded as a proof of the non-renormalisation theorem for the conformal anomaly \cite{Petkou:1999fv}:
\begin{align}
    \mathcal{A}_{+}-\mathcal{A}_{-} = \mathcal{A}_{k_T\neq 0}[\theta(-k_T) + \theta(+k_T)] = \mathcal{A}_{k_T\neq 0}
\end{align}
since a perturbative bulk field redefinition amounts to a renormalisation of the boundary source. In the scattering-amplitude framework, the calculation reduces to evaluating the residue of the integrand at $\eta=0$, hence it extends directly to the anomaly calculation via the identification
\begin{align}\label{Residue}
    a = \mathcal{A}_{k_T\neq 0} =i\text{Res}_{\eta = 0}\;\biggl[\frac{1}{\eta^D} F(ik\eta)e^{-ik_T\eta}\biggr].
\end{align}
Therefore, GEC--which by definition requires $\mathcal{A}_{k_T\neq0}=0$, so that the dS amplitude has support only when energy is conserved--corresponds precisely to the absence of a conformal anomaly.
 
Boundary locality of the anomaly is encoded in its analyticity in the kinematic variables. At four points, these variables are given by
\begin{align}
&k_1^2,\;k_2^2,\;k_3^2,\;k_4^2,\nonumber\\
&k_s^2 \equiv (k_1 + k_2)^{\alpha}(k_1 + k_2)_{\alpha},\;k_t^2\equiv (k_1 + k_3)^{\alpha}(k_1 + k_3)_{\alpha}.
\end{align}
To establish non-locality, one must enumerate all local counterterms. A concrete criterion was presented in \cite{Bzowski:2019kwd}: a divergent external configuration is holographically renormalisable only when
\begin{align} \label{Localcondition}
    d+\sum_{j=1}^4 \sigma_j (\Delta_{j} - \frac{d}{2}) = -2n,
\end{align}
for a non-negative integer $n$, where the signs $\sigma_j$ are either all minus $(----)$ or three minus and one plus $(---+)$. As we shall see in the next section, other sign configurations generate non-local anomalies and therefore impose strong constraints on the bulk interactions. Holographic consistency thus refers to the vanishing of bulk interaction couplings responsible for such non-local anomalies. We demonstrate that they can be cancelled completely by tuning the couplings among different bulk interactions. Because not every contribution to $a$ is non-local, the vanishing of the full anomaly required by GEC is stronger than the locality condition imposed by holographic consistency.
The conformal invariance implied by GEC agrees with known constructions of the dS/CFT correspondence \cite{Giombi:2016ejx,Brust:2016gjy,Brust:2016zns,Anninos:2011ui,Sezgin:2002rt}. The absence of logarithmic divergences was demonstrated explicitly in pure Einstein gravity \cite{Henningson:1998gx,deHaro:2000vlm}. It was also argued from a dS/CFT perspective in \cite{Anninos:2014lwa}, where the absence of a Weyl anomaly in an odd-dimensional CFT forces the absence of logarithms in the bulk. Here, we present a complementary bulk perspective: GEC, regarded as a vacuum-stability requirement in extended Poincar\'e dS spacetime, enforces the absence of a boundary anomaly.

Scalar fields dual to scalar operators with integer conformal dimensions, which are shown to be fixed by shift symmetries \cite{Bonifacio:2018zex,Farnsworth:2024yeh}, are also recently entertained in the context of scale separation in AdS vacua which contain moduli dual to those operators \cite{Conlon:2021cjk,Apers:2022tfm}. Integer dimensions appear to be a universal feature of scale-separated AdS vacua. The corresponding fields in dS are even more intriguing because they belong to the discrete series of unitary irreducible representations of the dS group \cite{Thomas1941,Dixmier1961}. In other words, they occupy isolated points in the landscape and cannot receive continuous quantum corrections. Together with GEC, or equivalently conformal invariance in the holographic dual, this property strongly constrains low-energy physics in dS.

To illustrate these points, we study four-point interactions involving three identical conformally coupled scalars and one discrete series scalar in $D=4$ dS and AdS spacetimes. We find that the GEC condition in dS is strictly stronger than holographic consistency in (A)dS. We perform both GEC and holographic bootstraps explicitly, finding unique interactions in both cases up to a specified derivative order and demonstrating that the GEC solutions form a strict subset of the holographically consistent solutions. Pertaining to the AdS$_4$ scale-separated vacua \cite{Bobev:2025yxp}, we found that a particular four-point interaction $\Delta_{1}=\Delta_{2}=\Delta_{3}=2, \Delta_4 = 8$ could be of interest for such vacua.

\section{Discrete series scalars with conformally coupled scalars}
In this section, we use \eqref{Residue} to compute the non-energy-conserving dS scattering amplitude $\mathcal{A}_{k_T\neq0}$ and the holographic anomaly $a$ for several four-point configurations involving three identical conformally coupled scalars and one massive discrete series scalar in $D=4$ dS spacetime. We then use these criteria to bootstrap the landscape of interactions. We begin with a single potential $\phi_{(a)}\phi_{(2)}^3$, where the subscript denotes the conformal dimension of the dual operator, and incrementally add derivative interactions until we obtain a solution to GEC. We proceed analogously for holographic consistency. Using the freedom to perform field redefinitions, we choose the higher-derivative interactions to have the convenient form $\Box^{m}[\phi_{(a)}\phi_{(2)}]\Box^{n}[\phi_{(2)}\phi_{(2)}]$, thereby avoiding longer chains of covariant derivatives. The use of holography to constrain effective actions in AdS was envisaged in \cite{El-Showk:2011yvt}; here, we put forth this idea in several simple illustrative examples.
\paragraph{$\bullet$ $\Delta_1 = \Delta_2 = \Delta_3 = 2, \Delta_4 = 4$}

In this case, GEC requires at least four derivatives. The Lagrangian ansatz is
\begin{align}
     \frac{\mathcal{L}_{4pt}}{\sqrt{-g}} = C_0 \phi_{(4)} \phi_{(2)}^3 + \frac{C_2}{H^4} \Box_{\text{bulk}} [\phi_{(4)} \phi_{(2)}] \Box_{\text{bulk}}[\phi_{(2)}\phi_{(2)}],
\end{align}
where the bulk Laplacian is defined with respect to the bulk (A)dS metric \eqref{metric}. The anomaly and the non-energy-conserving scattering amplitude follow from the residue formula \eqref{Residue}:
\begin{align}
    a = \mathcal{A}_{k_T\neq 0} = \frac{3}{4} (C_0 + 4 C_2)(k_1 + k_2 +k_3),
\end{align}
We defer this simple computation to the Supplemental Material. GEC therefore imposes
\begin{align}
    \text{GEC} (\text{anomaly free}): C_2 = -\frac{1}{4}C_0,
\end{align}
Holographic consistency imposes no constraint in this case because the anomaly can be renormalised by the semilocal counterterm
\begin{align}
    S^{\text{ct}} \propto \int d^3 x\, \bar{\phi}_{(4)} \bar{\phi}_{(2)}^2 \mathcal{O}_{(2)},
\end{align}
where $\bar{\phi}_{(\Delta)}$ is the source of dimension $d-\Delta$ that couples to the boundary operator $\mathcal{O}_{(\Delta)}$ of dimension $\Delta$. The holographic counterterm gives
\begin{align}
    a^{\text{ct}} &\propto (-i)^4 \frac{\delta}{\delta\bar{\phi}_{(4)}} \biggl(\frac{\delta}{\delta\bar{\phi}_{(2)}}\biggr)^3 \text{exp}[i (\int \bar\phi_{(2)}\mathcal{O}_{(2)} + S^{\text{ct}})] \nonumber\\
    &= \braket{\mathcal{O}_{(2)}\mathcal{O}_{(2)}} + \text{perms} \sim k_1 + k_2 + k_3,
\end{align}
given the renormalisable two-point function for discrete series scalars in even $D$ \cite{Blauvelt:2022wwa, Bzowski:2015pba}
\begin{align}
    \braket{\mathcal{O}_{(2)}(\vec{k})\mathcal{O}_{(2)}(-\vec{k})}\sim k^{2\Delta - d}|_{\Delta = 2,d=3} = k.
\end{align}
This renormalisability attests to the locality condition \eqref{Localcondition}. In this case, only the $(---+)$ condition is satisfied, and the anomaly can be renormalised completely without tuning the couplings.

\paragraph{$\bullet$ $\Delta_1 = \Delta_2 = \Delta_3 = 2, \Delta_4 = 5$}
For this configuration, GEC requires at least six derivatives. The Lagrangian ansatz is
\begin{align}
     \frac{\mathcal{L}_{4pt}}{\sqrt{-g}} = &C_0 \phi_{(5)} \phi_{(2)}^3 + \frac{C_2}{H^4} \Box_{\text{bulk}}[\phi_{(5)} \phi_{(2)}] \Box_{\text{bulk}}[\phi_{(2)}\phi_{(2)}] \nonumber\\
     + &\frac{C_3}{H^6} \Box_{\text{bulk}}[\phi_{(5)}\phi_{(2)}]\Box_{\text{bulk}}^2[\phi_{(2)}\phi_{(2)}].
\end{align}
The anomaly and the non-energy-conserving scattering amplitude are
\begin{align}
    a &=  \mathcal{A}_{k_T\neq 0} \nonumber\\
    &=- \frac{15}{8}i\biggl\{ [C_0 + 32(C_2 - 3C_3)](k_1^2 + k_2^2 + k_3^2- \frac{1}{5}k_4^2) \nonumber\\
    &+ 2 [C_0 + 16 (C_2 - 4 C_3)](k_1 k_2 + k_1 k_3 + k_2 k_3)\biggr\}.
\end{align}
GEC thus imposes
\begin{align}
    \text{GEC} (\text{anomaly free}): C_2 = \frac{1}{16} C_0,\;\; C_3 = \frac{1}{32} C_0.
\end{align}
For holographic consistency, the non-analyticity in two squared momenta $k_i k_j$ is non-holographically renormalisable, therefore the corresponding bulk interaction coupling must vanish. Holographic consistency thus imposes
\begin{align}
    \text{Holographic consistency}&\,(\text{local anomaly}):\nonumber\\
    C_0 = -16 (C_2 &- 4 C_3).
\end{align}
When the action is truncated at four derivatives, the interaction is unique. The remaining ultralocal anomaly can be removed by counterterms constructed purely from sources:
\begin{align}
    S^{\text{ct}}_{1} &\propto \int d^3x \;\Box_{\text{boundary}}\bar{\phi}_{(5)} \bar{\phi}_{(2)}^3,\nonumber\\
    S^{\text{ct}}_{2} &\propto \int d^3x\; \bar{\phi}_{(5)}\bar{\phi}_{(2)}^2\Box_{\text{boundary}}\bar{\phi}_{(2)},
\end{align}
where the box is defined with respect to the boundary metric, $\Box_{\text{boundary}}\equiv g^{\alpha\beta}\partial_\alpha\partial_\beta$. This result agrees with the locality condition \eqref{Localcondition}: both the $(----)$ and $(--++)$ conditions are satisfied, and the latter is not holographically renormalisable \cite{Bzowski:2019kwd}. Consequently, the interaction couplings must generally be tuned to cancel the non-local anomaly. We scamper through the $\Delta_4 = 6,7$ examples while explicating the $\Delta_4 = 8$ example which may be relevant to AdS scale-separated vacua.
\paragraph{$\bullet$ $ \Delta_1 = \Delta_2 = \Delta_3 = 2, \Delta_4 = 6$} 
The Lagrangian ansatz is given by
\begin{align}
     \frac{\mathcal{L}_{4pt}}{\sqrt{-g}} = &C_0 \phi_{(6)} \phi_{(2)}^3 + \frac{C_2}{H^4} \Box_{\text{bulk}}[\phi_{(6)} \phi_{(2)}] \Box_{\text{bulk}}[\phi_{(2)}\phi_{(2)}] \nonumber\\
     &+ \frac{C_3}{H^6} \Box_{\text{bulk}}[\phi_{(6)}\phi_{(2)}]\Box_{\text{bulk}}^2[\phi_{(2)}\phi_{(2)}]\nonumber\\
     &+ \frac{C_4}{H^8} \Box_{\text{bulk}}^2[\phi_{(6)}\phi_{(2)}]\Box_{\text{bulk}}^2[\phi_{(2)}\phi_{(2)}].
\end{align}
GEC and holographic consistency give
\begin{align}
&\text{GEC}~(\text{anomaly free}): \nonumber\\
&\qquad\qquad
C_2=\frac{1}{80}C_0,\;
C_3=\frac{1}{80}C_0,\;
C_4=\frac{1}{960}C_0.
\\
        &\text{Holographic consistency}\,(\text{local anomaly}):\nonumber\\
    &\qquad \qquad \qquad C_0 = -48 C_2 + 192 C_3 - 768 C_4.
\end{align}

\paragraph{$\bullet$ $ \Delta_1 = \Delta_2 = \Delta_3 = 2, \Delta_4 = 7$} 
The Lagrangian ansatz is given by
\begin{align}
     &\frac{\mathcal{L}_{4pt}}{\sqrt{-g}} = C_0 \phi_{(7)} \phi_{(2)}^3 + \frac{C_2}{H^4} \Box_{\text{bulk}}[\phi_{(7)} \phi_{(2)}] \Box_{\text{bulk}}[\phi_{(2)}\phi_{(2)}] \nonumber\\
     &+ \frac{C_3}{H^6} \Box_{\text{bulk}}[\phi_{(7)}\phi_{(2)}]\Box_{\text{bulk}}^2[\phi_{(2)}\phi_{(2)}]\nonumber\\
     &+ \frac{C_4}{H^8} \Box_{\text{bulk}}^2[\phi_{(7)}\phi_{(2)}]\Box_{\text{bulk}}^2[\phi_{(2)}\phi_{(2)}]\nonumber\\
     &+ \frac{C_5}{H^{10}} \Box_{\text{bulk}}^2[\phi_{(7)}\phi_{(2)}]\Box_{\text{bulk}}^3[\phi_{(2)}\phi_{(2)}].
\end{align}
GEC and holographic consistency give
\begin{align}
&\text{GEC}~(\text{anomaly free}): \nonumber\\
&C_2 = \frac{17}{3960} C_0,\quad
C_3 = \frac{19}{2640} C_0,\quad
C_4 = \frac{1}{1056} C_0,\quad\nonumber\\
&C_5 = \frac{1}{31680} C_0.\\
        &\text{Holographic consistency}\,(\text{local anomaly}):\nonumber\\
    &\qquad C_0 = - 364 C_2 + 6232 C_3 - 108976 C_4 + 1929568 C_5.
\end{align}

\paragraph{$\bullet$ $ \Delta_1 = \Delta_2 = \Delta_3 = 2, \Delta_4 = 8$}
In this case, we include operators with up to twelve derivatives modulo field basis change \footnote{There are two independent twelve-derivative operators, however, they are degenerate at the level of the conformal anomaly. Hence, up to twelve derivatives we are left with one un-fixed interaction.}
\begin{align}
     &\frac{\mathcal{L}_{4pt}}{\sqrt{-g}} = C_0 \phi_{(8)} \phi_{(2)}^3 + \frac{C_2}{H^4} \Box_{\text{bulk}}[\phi_{(8)} \phi_{(2)}] \Box_{\text{bulk}}[\phi_{(2)}\phi_{(2)}] \nonumber\\
     &+ \frac{C_3}{H^6} \Box_{\text{bulk}}[\phi_{(8)}\phi_{(2)}]\Box_{\text{bulk}}^2[\phi_{(2)}\phi_{(2)}] \nonumber\\
     &+ \frac{C_4}{H^8} \Box_{\text{bulk}}^2[\phi_{(8)}\phi_{(2)}]\Box_{\text{bulk}}^2[\phi_{(2)}\phi_{(2)}]\nonumber\\
     &+ \frac{C_5}{H^{10}} \Box_{\text{bulk}}^2[\phi_{(8)}\phi_{(2)}]\Box_{\text{bulk}}^3[\phi_{(2)}\phi_{(2)}] \nonumber\\
     &+ \frac{C_6}{H^{12}}\Box_{\text{bulk}}^3[\phi_{(8)}\phi_{(2)}]\Box_{\text{bulk}}^3[\phi_{(2)}\phi_{(2)}].\label{FourptLag}
\end{align}
Once again, there is a unique GEC solution:
\begin{align}
&\text{GEC(anomaly free)}:\\
&C_2=\frac{37}{21420}C_0,\qquad C_3=\frac{163}{34272}C_0,\qquad C_4=\frac{89}{114240}C_0,\nonumber\\ 
&C_5=\frac{29}{685440}C_0,\qquad C_6=\frac{1}{1370880}C_0.
\end{align}
As in the preceding cases, the non-analytic contributions $a\supset k_1k_2k_3k_4^2,\,k_1k_2k_3^3+\text{perms}$ cannot be generated by any local counterterm. Cancelling the non-local anomaly therefore requires the coupling relation
\begin{align}
   &\text{Holographic consistency} (\text{local anomaly}): \\
   &C_0 = -532 C_2 + 9256 C_3 - 165328 C_4\nonumber\\
   &\qquad + 2970784 C_5 - 53453632 C_6,\label{ScaleSeparation}
\end{align}
When the higher-derivative interactions are set to zero, this relation pins down a unique interaction up to four derivatives. We present the anomaly at this order in the Supplemental Material. The relevant semilocal counterterms are
\begin{align}
    &S^{\text{ct}}\sim \int d^3x\;\partial^{p_1}\bar \phi_{(8)}\partial^{p_2}\bar{\phi}_{(2)}\partial^{p_3}\bar{\phi}_{(2)} \partial^{p_4} \mathcal{O}_{(2)},
\end{align}
where one sums over all partitions of $p_1+p_2+p_3+p_4=4$ and all possible contractions with respect to the boundary metric. The result agrees with the locality condition \eqref{Localcondition}: both the $(---+)$ and $(-+++)$ conditions are satisfied, and the latter is not holographically renormalisable. The couplings must therefore generally be tuned to cancel the non-local anomaly.

This case is particularly relevant to scale separation in AdS because such an interaction could occur in non-supersymmetric AdS$_4$ flux vacua \cite{Conlon:2021cjk, Bobev:2025yxp}; see the Supplemental Material.

In all the cases above, GEC imposes strictly stronger constraints than holographic consistency. It would be interesting to explore whether there exist generalised shift symmetries underlie these unique GEC solutions.

\section{Discussion}
Through concrete new examples, we have demonstrated a precise relation between GEC in dS scattering amplitudes and holographic consistency for contact interactions. We have discovered novel, unique interactions involving conformally coupled scalars and discrete series scalars, and we anticipate that underlying shift symmetries may account for both holographic consistency and GEC.

We have established this identification for contact interactions; however, a complete understanding at full tree level remains elusive. It is tempting to identify the $1/z$ coefficient of the Lagrangian evaluated on the classical saddle directly with both the anomaly and the non-energy-conserving amplitude. This identification fails once exchange diagrams are included, because $\log z$ terms can be generated at higher orders in the source expansion. Moreover, the meaning of ``conformal anomaly'' becomes obscure for a generic holographic CFT at a given perturbative order, whereas the non-energy-conserving term remains well defined to all orders in the classical action.

In future work, we aim to demonstrate that the (A)dS scattering amplitude itself can be interpreted as a full-fledged, exactly conformally invariant holographic correlator without additional anomalous contributions to all orders at bulk tree level.

\section{Acknowledgements}

The author would like to thank Paul Saffin, Kosta Skenderis, and David Stefanyszyn for enlightening discussions. The author would also like to thank Jasper Eon and Bartosz Pyszkowski for pointing out several typos in the first pre-print. 
ZD is supported by Simons Investigator award SFI-MPS-SIP-00017355.

\bibliographystyle{apsrev4-1}
\bibliography{RefDBI}

\appendix
\onecolumngrid
{\begin{center}\bf \Large{Supplemental Material}\end{center} }
\setlength{\abovedisplayskip}{6pt}
\setlength{\abovedisplayshortskip}{3pt}
\setlength{\belowdisplayskip}{6pt}
\setlength{\belowdisplayshortskip}{3pt}

\section{Contour deformation}

In this section, we elucidate the contour deformation that we have alluded to in the main text. We begin with the integral
\begin{align}
   I = \int_{-\infty}^{+\infty}\frac{d\eta}{\eta^D} F(ik\eta) e^{-ik_T\eta},
\end{align}
which can be evaluated by decomposing it into integrals of the form
\begin{align}
    I_p \equiv \int_{-\infty}^{+\infty}d\eta\;\eta^{p} e^{-i k_T \eta},\;\;p\leq-1,
\end{align}
where we focus on the terms that diverge at $\eta=0$.
\begin{figure}
    \centering
\begin{center}
    \centering
    \begin{tikzpicture}[scale = 2]
        \draw[black, ->] (-1.6,0) -- (1.6,0) coordinate (xaxis);
        \draw[black, ->] (0,-1.6) -- (0,1.7) coordinate (yaxis);
        \node at (1.85, 0) {$\Re\eta$};
        \node at (0, 1.85) {$\Im\eta$};
		
        \draw[pyred, fill = pyred] (0,0) circle (.03cm);
        
        \node at (-0.8, -0.25) {\textcolor{pypurple}{$\mathcal C_1$}};
        \node at (0.8, -0.25) {\textcolor{pypurple}{$\mathcal C_3$}};
        \node at (-0.24, 0.24) {\textcolor{pypurple}{$\mathcal C_2$}};
        \node at (1.2, -1.2) {\textcolor{pyblue}{$\mathcal C_4$}};
        \node at (1.2, 1.3) {\textcolor{pyred}{$\mathcal C_4'$}};

        \path[pyred, draw, line width = 1.0pt] (0.25,0.55) -- (0.6,0.55);
        \path[pyblue, draw, line width = 1.0pt] (0.25,-0.55) -- (0.6,-0.55);
        \node at (0.9, 0.55) {$k_T\leq 0$};
        \node at (0.9, -0.55) {$k_T\geq 0$};

        \path[pypurple, draw, line width = 0.8pt, postaction = decorate, 
		decoration={markings,
			mark=at position 0.5 with {\arrow[line width=1pt]{>}}}] (-1.5, 0) -- (-0.15,0);
            \path[pypurple, draw, line width = 0.8pt, postaction = decorate, 
		decoration={markings,
			mark=at position 0.5 with {\arrow[line width=1pt]{>}}}] (0.15,0) -- (1.5,0);
	\path[pyred, draw, dashed, line width = 0.8pt, postaction = decorate, 
		decoration={markings,
			mark=at position 0.25 with {\arrow[line width=1pt]{>}}}] (1.5,0.) arc (-3.81407:180+3.81407:1.50333);

	\path[pyblue, draw, dashed, line width = 0.8pt, postaction = decorate, 
		decoration={markings,
			mark=at position 0.25 with {\arrow[line width=1pt]{>}}}] (1.5,0) arc (-3.81407:-180+3.81407:1.50333);
	\path[pypurple, draw,  line width = 0.8pt, postaction = decorate, 
		decoration={markings,
			mark=at position 0.25 with {\arrow[line width=1pt]{>}}}] (-0.15,0) arc (-180:0:0.15);
	\end{tikzpicture}
\end{center}
    \caption{The time-integration contour for a dS scattering amplitude in an eternally energy-creating universe. The contour is chosen so that the integral over the large arc vanishes for $k_T\neq0$, while the scattering amplitude is non-vanishing only when $k_T=k_{\text{in}}-k_{\text{out}}\leq0$.}
    \label{EnergyCreation}
\end{figure}

\begin{figure}
    \centering
\begin{center}
    \centering
    \begin{tikzpicture}[scale = 2]
        \draw[black, ->] (-1.6,0) -- (1.6,0) coordinate (xaxis);
        \draw[black, ->] (0,-1.6) -- (0,1.7) coordinate (yaxis);
        \node at (1.85, 0) {$\Re\eta$};
        \node at (0, 1.85) {$\Im\eta$};
		
        \draw[pyred, fill = pyred] (0,0) circle (.03cm);
        
        \node at (-0.8, -0.25) {\textcolor{pypurple}{$\mathcal C_1$}};
        \node at (0.8, -0.25) {\textcolor{pypurple}{$\mathcal C_3$}};
        \node at (-0.24, 0.24) {\textcolor{pypurple}{$\mathcal C_2$}};
        \node at (1.2, -1.2) {\textcolor{pyblue}{$\mathcal C_4$}};
        \node at (1.2, 1.3) {\textcolor{pyred}{$\mathcal C_4'$}};

        \path[pyred, draw, line width = 1.0pt] (0.25,0.55) -- (0.6,0.55);
        \path[pyblue, draw, line width = 1.0pt] (0.25,-0.55) -- (0.6,-0.55);
        \node at (0.9, 0.55) {$k_T\leq 0$};
        \node at (0.9, -0.55) {$k_T\geq 0$};

        \path[pypurple, draw, line width = 0.8pt, postaction = decorate, 
		decoration={markings,
			mark=at position 0.5 with {\arrow[line width=1pt]{>}}}] (-1.5, 0) -- (-0.15,0);
            \path[pypurple, draw, line width = 0.8pt, postaction = decorate, 
		decoration={markings,
			mark=at position 0.5 with {\arrow[line width=1pt]{>}}}] (0.15,0) -- (1.5,0);
	\path[pyred, draw, dashed, line width = 0.8pt, postaction = decorate, 
		decoration={markings,
			mark=at position 0.25 with {\arrow[line width=1pt]{>}}}] (1.5,0.) arc (-3.81407:180+3.81407:1.50333);

	\path[pyblue, draw, dashed, line width = 0.8pt, postaction = decorate, 
		decoration={markings,
			mark=at position 0.25 with {\arrow[line width=1pt]{>}}}] (1.5,0) arc (-3.81407:-180+3.81407:1.50333);
	\path[pypurple, draw,  line width = 0.8pt, postaction = decorate, 
		decoration={markings,
			mark=at position 0.25 with {\arrow[line width=1pt]{>}}}] (-0.15,0) arc (180:0:0.15);
	\end{tikzpicture}
\end{center}
    \caption{The time-integration contour for a dS scattering amplitude in an eternally energy-annihilating universe. The contour is chosen so that the integral over the large arc vanishes for $k_T\neq0$, while the scattering amplitude is non-vanishing only when $k_T=k_{\text{in}}-k_{\text{out}}\geq0$.}
    \label{EnergyAnnihilation}
\end{figure}

	The analytic structures plotted in Figures \ref{EnergyCreation} and \ref{EnergyAnnihilation} apply to discrete series scalars in even spacetime dimensions, for which $\nu$ is a half-integer. For on-shell interactions, the integrand is analytic everywhere except at the pole at the conformal boundary, $\eta=0$.

In the complex $\eta$ plane, we aim to evaluate the integral along the real axis, which is barricaded by the pole at the origin. We therefore deform the contour above or below the pole along an infinitesimal semicircle, represented by the purple path $\textcolor{pypurple}{\mathcal{C}_1}\rightarrow\textcolor{pypurple}{\mathcal{C}_2}\rightarrow\textcolor{pypurple}{\mathcal{C}_3}$, here $\mathcal{C}_2$ identifies with $\mathcal{C}_\epsilon$ in the main text. The two deformations correspond to different ''boundary'' conditions at the conformal boundary.

The purple line integral can then be evaluated using the residue theorem by closing the contour with a large arc at infinity, represented by the semicircles $\textcolor{pyred}{\mathcal{C}_4'}$ and $\textcolor{pyblue}{\mathcal{C}_4}$. We close the contour in whichever half-plane makes the large-arc contribution vanish. A phase $e^{-ik_T\eta}$ can always be factored out of the integrand. Consequently, the integrand is exponentially damped along the upper-half-plane arc $\textcolor{pyred}{\mathcal{C}_4'}$ when $k_T<0$, and along the lower-half-plane arc $\textcolor{pyblue}{\mathcal{C}_4}$ when $k_T>0$.

Focusing on Figure \ref{EnergyCreation}, when $k_T<0$, we close the contour in the upper half-plane. The full contour $\textcolor{pypurple}{\mathcal{C}_1}\rightarrow\textcolor{pypurple}{\mathcal{C}_2}\rightarrow\textcolor{pypurple}{\mathcal{C}_3}\rightarrow\textcolor{pyred}{\mathcal{C}_4'}$ encircles the pole at the origin. The residue theorem therefore gives
\begin{align}
    I_{\textcolor{pypurple}{\mathcal{C}_1}\rightarrow\textcolor{pypurple}{\mathcal{C}_2}\rightarrow\textcolor{pypurple}{\mathcal{C}_3}}(k_T<0) = -I_{\textcolor{pyred}{\mathcal{C}_4'}} + 2\pi i \text{Res}_{\eta = 0} = 2\pi i \text{Res}_{\eta = 0}.
\end{align}
When $k_T>0$, we close the contour in the lower half-plane. In this case, the contour encloses no poles, so the integral vanishes. When $k_T=0$, the large-arc contribution does not vanish for $p=-1$. It can nevertheless be evaluated explicitly:
\begin{align}
    \int_{\textcolor{pyred}{\mathcal{C}_4'}} \frac{d\eta}{\eta} = \int_{0}^{\pi} \frac{d e^{i\theta}}{e^{i\theta}} = i\pi.
\end{align}
Because the coefficient multiplying this integral is precisely the residue, the real-line integral is
\begin{align}
    I_{\textcolor{pypurple}{\mathcal{C}_1}\rightarrow\textcolor{pypurple}{\mathcal{C}_2}\rightarrow\textcolor{pypurple}{\mathcal{C}_3}}(k_T = 0) = -I_{\textcolor{pyred}{\mathcal{C}_4'}} + 2\pi i \text{Res}_{\eta = 0} = \pi i \text{Res}_{\eta = 0}.
\end{align}
These three cases can be encapsulated by a single step function $\theta(-k_T)$, where
\begin{align}
\theta(x) = 
\begin{cases}
    1,\qquad x>0\\
    \frac{1}{2},\qquad x=0\\
    0,\qquad x<0
\end{cases}
\end{align}
As a consistency check, we can instead close the contour along $\textcolor{pyblue}{\mathcal{C}_4}$ when $k_T=0$. In this case, the purple line integral is determined by the arc integral:
\begin{align}
    I_{\textcolor{pypurple}{\mathcal{C}_1}\rightarrow\textcolor{pypurple}{\mathcal{C}_2}\rightarrow\textcolor{pypurple}{\mathcal{C}_3}}(k_T = 0) = -I_{\textcolor{pyblue}{\mathcal{C}_4}} = - \text{Res}_{\eta = 0}\int_{0}^{-\pi} \frac{d e^{i\theta}}{e^{i\theta}} = \pi i \text{Res}_{\eta = 0}, 
\end{align}
which agrees with the previous calculation. The analysis of Figure \ref{EnergyAnnihilation} is nearly identical, except that both the argument of the step function and the overall coefficient change sign. At $k_T=0$, the analytic part of the integrand gives additional contributions to the energy-conserving component of the amplitude:
\begin{align}
    \mathcal{A}\supset \sum_{m}\mathcal{A}^{(m)}\delta^{(m)}(k_T).
\end{align}
\section{Computational details}
We next illustrate the computation of the holographic anomaly. Consider the four-point interaction of $\Delta_1 = \Delta_2 =\Delta_3 = 2,\,\Delta_4 = 4$
\[
\frac{\mathcal{L}_{4pt}}{\sqrt{-g}} = C_0 \phi_{(4)} \phi_{(2)}^3 + \frac{C_2}{H^4} \Box_{\text{bulk}}[ \phi_{(4)} \phi_{(2)}] \Box_{\text{bulk}}[\phi_{(2)}\phi_{(2)}],
\]
as an example. The holographic anomaly is
    \begin{align}
    a &= \mathcal{A}_{k_T\neq 0} \nonumber\\
    &= i \text{Res}_{\eta = 0}\biggl\{ \frac{1}{\eta^4}\biggl[C_0 \phi_{(2)}(k_1,\eta)\phi_{(2)}(k_2,\eta)\phi_{(2)}(k_3,\eta)\phi_{(4)}(k_4,\eta) \nonumber\\
    &+ C_2 \eta^4[\partial_{\eta}^2 -  \frac{2}{\eta}\partial_{\eta} + (\vec{k}_1 + \vec{k}_2)^2][\phi_{(2)}(k_1,\eta)\phi_{(2)}(k_2,\eta)][\partial_{\eta}^2 -  \frac{2}{\eta}\partial_{\eta} + (\vec{k}_3 + \vec{k}_4)^2][\phi_{(2)}(k_3,\eta)\phi_{(4)}(k_4,\eta)] + \text{perms} \biggr]\biggr\} \nonumber\\
    &=\frac{3}{4} (C_0 + 4 C_2)(k_1 + k_2 +k_3),
\end{align}
where the bulk-to-boundary propagator is given by \eqref{BulktoBoundary}, and momentum conservation has been imposed to eliminate all dependence on $\vec{k}_4$. Specifically, the bulk-to-boundary propagators for the $\Delta=2$ and $\Delta=4$ scalars are
\begin{align}
    \phi_{(2)}(k,\eta) = \frac{\eta}{\sqrt{2}} e^{-ik\eta},\;\;\phi_{(4)}(k,\eta) = \frac{1}{\sqrt{2}\eta}[-3i + k\eta(3+ ik\eta)] e^{-i  k \eta}.
\end{align}
\section{$\Delta_1 =\Delta_2 =\Delta_3 = 2, \Delta_4 = 8$ anomaly}
Motivated by scale-separated vacua, we present the anomaly and the non-energy-conserving amplitude for the ansatz \eqref{FourptLag}, truncated at four derivatives:
\[
\begin{aligned}
a &= \mathcal{A}_{k_T\neq 0}\nonumber\\
&=\frac{21}{32}\biggl\{
33(C_0+1652C_2)k_1^5+33(C_0+932C_2)k_2^5+165(C_0+676C_2)k_2^4k_3
+165(C_0+1012C_2)k_1^4(k_2+k_3)\\
&\quad+30k_1^3\biggl[
11(C_0+788C_2)k_2^2+22(C_0+532C_2)k_2k_3+11(C_0+788C_2)k_3^2
-(C_0-28C_2)k_4^2-1056C_2(k_s^2+k_t^2)
\biggr]\\
&\quad+30k_2^3\biggl[
11(C_0+620C_2)k_3^2-(C_0+716C_2)k_4^2-264C_2(k_s^2-2k_t^2)
\biggr]\\
&\quad+30k_2^2k_3\biggl[
11(C_0+620C_2)k_3^2-3(C_0+844C_2)k_4^2+792C_2(k_s^2+2k_t^2)
\biggr]\\
&\quad+k_3\biggl[
33(C_0+932C_2)k_3^4+5(C_0+1156C_2)k_4^4
-720C_2k_4^2(13k_s^2+11k_t^2)+7920C_2(k_s^4+k_s^2k_t^2+k_t^4)\\
&\qquad\qquad
-30k_3^2\biggl(
(C_0+716C_2)k_4^2+264C_2(-2k_s^2+k_t^2)
\biggr)
\biggr]\\
&\quad+5k_2\biggl[
33(C_0+676C_2)k_3^4+(C_0+1156C_2)k_4^4
-144C_2k_4^2(11k_s^2+13k_t^2)+1584C_2(k_s^4+k_s^2k_t^2+k_t^4)\\
&\qquad\qquad
-18k_3^2\biggl(
(C_0+844C_2)k_4^2-264C_2(2k_s^2+k_t^2)
\biggr)
\biggr]\\
&\quad+30k_1^2\biggl[
11(C_0+788C_2)k_2^3+33(C_0+436C_2)k_2^2k_3+11(C_0+788C_2)k_3^3\\
&\qquad\qquad
+3k_2\biggl(
11(C_0+436C_2)k_3^2-(C_0+228C_2)k_4^2-352C_2(2k_s^2+k_t^2)
\biggr)\\
&\qquad\qquad
-3k_3\biggl(
(C_0+228C_2)k_4^2+352C_2(k_s^2+2k_t^2)
\biggr)
\biggr]\\
&\quad+5k_1\biggl[
33(C_0+868C_2)k_2^4+132(C_0+532C_2)k_2^3k_3+33(C_0+868C_2)k_3^4
+(C_0+868C_2)k_4^4\\
&\qquad\qquad
+12(C_0+532C_2)k_2k_3(11k_3^2-3k_4^2)
-1296C_2k_4^2k_s^2+1584C_2k_s^4
+144C_2(-9k_4^2+11k_s^2)k_t^2+1584C_2k_t^4\\
&\qquad\qquad
+18k_2^2\biggl(
11(C_0+444C_2)k_3^2-(C_0+508C_2)k_4^2+88C_2(-5k_s^2+2k_t^2)
\biggr)\\
&\qquad\qquad
-18k_3^2\biggl(
(C_0+508C_2)k_4^2+88C_2(-2k_s^2+5k_t^2)
\biggr)
\biggr]
\biggr\}.
\end{aligned}
\]
This anomaly contains two non-local terms:
\begin{align}
    a \supset k_1 k_2 k_3 k_4^2, k_1 k_2 k_3^3 + \text{perms}.
\end{align}
These terms cannot be holographically renormalised and share the coefficient $C_0+532C_2$. We have verified explicitly that all the conformal anomalies computed above satisfy the exact conformal Ward identities.
\section{Scale-separated vacua}
The K\"ahler-sector spectrum of the vacua with sign $s=(1,1,1)$ is \cite{Conlon:2021cjk}
\begin{align}
    \Delta_{\varphi} = \{10,6,6,6\},\;\Delta_{a} = \{2,8,8,8\},
\end{align}
where the $\varphi$ are scalars and the $a$ are pseudoscalars. The pseudoscalars enjoy a $Z_2$ symmetry under the simultaneous sign flip $a_i\rightarrow-a_i$ for every $i$. We disregard the complex moduli because they decouple from the K\"ahler sector \cite{Apers:2022tfm}. The four-point process $a_{\Delta=8},a_{\Delta=2}\rightarrow a_{\Delta=2},a_{\Delta=2}$ receives no on-shell exchange contribution, because all on-shell cubic vertices are either absent or removable by field redefinitions. The exchanged particle cannot be an $a$, since the $Z_2$ symmetry forbids the interaction $a_i a_j a_k$. If the exchanged particle is a $\varphi$, the vertices $a_{\Delta=8}a_{\Delta=2}\varphi_{\Delta=6,10}$ are extremal and therefore can be removed by three-point holographic consistency \cite{Bobev:2025yxp}. If the theory is holographically consistent, \eqref{ScaleSeparation} therefore provides a prediction for this particular on-shell interaction.

\end{document}